\documentclass[pra,showpacs,floatfix,amsmath,amsfonts,twocolumn]{revtex4}
\usepackage{graphics}
\usepackage{epsfig}
\usepackage{color}

\newcommand{\cH}{{\cal H}}
\newcommand{\tx}{\tilde{x}}

\newcommand{\bPsi}{\mathbf{\Psi }}
\newcommand{\bpsi}{ \mbox{\boldmath$\psi$\unboldmath}  }
\newcommand{\bvarphi}{ \mbox{\boldmath$\varphi$\unboldmath}  }

\newcommand{\bp}{\mathbf{p }}
\newcommand{\bP}{\mathbf{P }}
\newcommand{\bU}{\mathbf{U }}
\newcommand{\bV}{\mathbf{V }}

\begin{document}

\title{Gap solitons in the spin-orbit coupled Bose-Einstein condensates}

\author{Yaroslav V. Kartashov$^{1,2}$, Vladimir V. Konotop$^{3}$, and Fatkhulla Kh. Abdullaev$^{4}$}

\affiliation{$^{1}$ ICFO-Institut de Ciencies Fotoniques, and Universitat Politecnica de Catalunya, 08860 Castelldefels (Barcelona), Spain
\\
$^{2}$
Institute of Spectroscopy, Russian Academy of Sciences, Troitsk, Moscow Region, 142190, Russia
\\
$^{3}$Centro de F\'isica Te\'{o}rica e Computacional and Departamento de
F\'isica, Faculdade de Ci\^encias, Universidade de Lisboa, Avenida Professor
Gama Pinto 2, Lisboa 1649-003, Portugal \\
$^{4}$ Instituto de F\'isica Te\'orica, Universidade Estadual Paulista, 01140-070, Sao Paulo, Brasil}
\date{\today}

\begin{abstract}

We report a diversity of stable gap solitons in a spin-orbit coupled Bose-Einstein condensate subject to a spatially periodic Zeeman field. It is shown that the solitons, can be classified by the main physical symmetries they obey, i.e. symmetries with respect to parity (P), time (T), and  internal degree of freedom, i.e. spin, (C) inversions. The conventional gap and gap-stripe solitons are obtained in lattices with different parameters. It is shown that solitons of the same type but obeying different symmetries can exist in the same lattice at different spatial locations. PT and CPT symmetric solitons have anti-ferromagnetic structure and are characterized respectively by nonzero and zero total magnetizations.

\end{abstract}
\maketitle


The progressively growing interest in the physics of mixtures of  spinor Bose-Einstein condensates (BECs) and in particular in spin-orbit (SO) coupled BECs (SO-BECs)~\cite{Nature,experiments}, is motivated on the one hand  by their fundamental importance for the atomic physics, and on the other hand by the rich possibilities they offer for emulating synthetic electric and magnetic fields \textcolor{black}{in solids~\cite{Spielman}}.  The last property allows one using neutral atoms
to simulate numerous condensed matter phenomena~\cite{emulator} in a tunable way, what would be impossible in direct condensed matter experiments. \textcolor{black}{In other words, a spinor BEC is a  promising candidate for implementation of a quantum simulator, which due to the technologies available nowadays was already implemented using cold atoms~\cite{simul1}, ions~\cite{simul2}, or photonics~\cite{simul3}. In particular, we mention a recent experiment~\cite{optics} with coupled (linear) arrays of optical waveguides simulating a discrete version of SO-coupling which in the present letter is considered for the matter waves.}

Meanwhile, the physics of BECs is characterized by two essential factors, \textcolor{black}{which are not typical for the physics which is aimed to be simulated.} First, a BEC is a nonlinear system, with the nonlinearity stemming from the two-body interactions~\cite{books}. Moreover, by changing the configuration of the system (i.e. of the laser beams) one can create effective nonlinear interactions of different types~\cite{nonlinearity}.
Respectively, such nonlinear objects as skyrmions~\cite{skyrmions}, solitons~\cite{AFKP,XZW}, anti-ferromagnetic structures and symmetry breaking~\cite{ZDKM}, have been recently reported for SO-BECs. The second feature, is that a (quasi-) stationary state of the condensate requires the presence of the external potential. This, in particular, imposes constraints on the lower bound of the kinetic energy, what is particularly relevant for SO-BECs (see e.g. \cite{ZDKM}).
What concerns the external potentials, since the very first experiments with cold atoms loaded in optical lattices~\cite{Kasevich} it was widely recognized that lattices (i.e. periodic potentials) are particularly efficient in manipulating BECs~\cite{MO}. Nowadays, the nonlinear properties of the atomic BECs held in optical lattices, described in the mean-field approximation are very well studied~\cite{rev_mean} (see also~\cite{mixtures} for a brief review on BEC mixtures in optical lattices). It is therefore natural that studies of SO-BECs in periodic potentials were already initiated. Manipulations with the band-gap structure~\cite{flat-band} and vortex lattice states~\cite{Sakaguchi} supported by SO coupling and optical lattices have being reported.

On the other hand a few years ago there have been initiated studies of the properties of electronic gas with Rashba~\cite{Rashba} and Dresselhaus~\cite{Dresselhaus} SO coupling
 in periodic structures. Those studies included the effect of dimensionality and features of the band-gap structures of a gas in a superimposed periodic and arbitrary confining potentials~\cite{SBG}, ballistic transport of electrons in periodic magnetic field~\cite{LDXZM,XLC}, switching controlled by SO coupling~\cite{GE}, and  quantum wire superlattices~\cite{TELE}. It was shown that SO coupling induced by an external periodic  field leads to the appearance of new gaps, making the electronic system tunable even in the condensed matter settings.

Even more recently, the SO coupling was emulated in experiments with light propagating in arrays optical waveguides~\cite{optics}, thus opening a root towards new physical systems where the interplay of SO coupling and nonlinearity (if waveguides are of the Kerr type) can result in novel optical phenomena.

Bearing in mind the described findings \textcolor{black}{as well as experimental availability of the spatially {\em periodic} Zeeman field~\cite{PeriodZeeman}} it is natural to address the effect of the lattice induced by such a field on the {\em nonlinear} properties of SO-BECs.  \textcolor{black}{In this Letter we, for the first time, perform thorough analysis of unusual properties,  symmetries, and stability of gap solitons in spinor condensates with periodically coupled components. We show that the periodic magnetic fields applied to a SO-BEC can support symmetric and asymmetric lattice solitons}, classified with respect to their parity (P), time (T), and pseudo-charge (C) symmetries. The chemical potentials of the solitons originate from different gaps in the spectrum of the magnetic lattice for both attractive and repulsive interactions.

We address one-dimensional (1D) mixture of a two-component SO-BEC which  is described by a spinor   ${\mathbf{\Psi }}=$col$(\Psi
^{(1)},\Psi^{(2)})$. In particular, for a tripod-type atomic scheme, like the one considered in~\cite{Edmonds},  the components $\Psi^{(1)}$ and $\Psi^{(2)}$ are the wavefunctions of the two dark states emulating (pseudo-)spins $|\!\!\uparrow \rangle $  and $|\!\!\downarrow \rangle $. Adopting the units where $\hbar =m=1$, and taking into account that the difference between spin-independent and spin-dependent two-body interactions can be made as small as necessary (in~\cite{Nature} the difference was of order of 1\% of the interaction magnitude) we explore the Hamiltonian of the system in the form
$H= \int_{-\infty
}^{\infty } \mathbf{\Psi }^{\dag }\left(\mathcal{H} +\frac{g}{2}\mathbf{\Psi }^\dag\mathbf{\Psi }\right)\mathbf{\Psi }dx$ where the linear Hamiltonian
$
\mathcal{H}= -\frac 12 \partial_x ^{2}+\frac 12 \Omega (x)
\sigma _{3} + i\kappa \sigma _{1}\partial_x
$
accounts for the periodically modulated Zeeman field $\Omega(x)$
and for SO coupling with the strength $\kappa$ (which depends on the angle between optical beams and can be tuned~\cite{review}), $g=$sign$\,a_s$ characterizes the of two-body interactions with the scattering length $a_s$,  and $\sigma_{1,2,3}$   are the Pauli matrices. The nonlinear dynamics of the condensate is then governed by the coupled Gross-Pitaevskii equations (GPEs)
\begin{equation}
i\bPsi_t = -\frac{1}{2}\bPsi_{xx} + \frac{\sigma_3}{2}\Omega(x)\bPsi  +
i\kappa\sigma_1 \bPsi_x
+ g(\bPsi^\dag\bPsi)\bPsi.
\label{GPE}
\end{equation}

We are interested in the stationary solutions having the form $\bPsi(x,t)=e^{-i\mu t}\bpsi(x)$, where $\mu$ is a chemical potential and $\bpsi(x)$ solves the stationary coupled GPEs.
For the sake of definiteness, we consider a particular form of the Zeeman field
\begin{eqnarray}
\label{omega}
\Omega(x)=2\left[\Delta+\delta\cos(2 x)\right],
\end{eqnarray}
where $\Delta\geq 0$ and $\delta\geq 0$ are the average and the modulation amplitude  of the field, respectively, and without loss of generality the period is taken to be $\pi$. Nevertheless we emphasize that the approach  developed below is valid for more generic forms of Zeeman fields obeying the symmetries specified below.

First we take into account that $\Omega(x)$ is an even function, $\Omega(x)=\Omega(-x)$ and  obtain that if $\bpsi(x)$ is a solution of Eq.~(\ref{GPE}) then the spinor
$
PT\bpsi(x)=\bar{\bpsi}(-x)
$
(an overbar stands for the complex conjugation)  defined by the parity and time inversions~\cite{com}, is a solutions of the stationary GPEs, as well.
At $\Delta=0$ the field (\ref{omega}) allows for additional symmetry reductions. The first one is readily seen in the shifted spatial variable $\tx=x+\pi/4$. Indeed, since $\Omega(\tx)=-\Omega(-\tx)$, if $\bpsi(\tx)$ is a solution of (\ref{GPE}) then the function obtained through the reduction $CPT  {\bpsi}(\tx)=\sigma_1\bar{\bpsi}(-\tx)$, is also a solution ($C$ by analogy with the charge symmetry~\cite{LL} indicates the symmetry related to the inversion of the internal degree of freedom, which is the pseudo-spin in our case).
Finally, due to the property $\Omega(x)=-\Omega(x+\pi/2)$ one more symmetry stems from the possibility of flipping spin components in the $\pi/2$ shifted lattice:  $S\bpsi (x)=\sigma_1\bpsi(x+\pi/2)$ is a solution of the stationary GPEs.

Bearing in mind the symmetries of the GPEs it is natural to use them for classifying the  solutions, i.e. to identify solutions which would be the eigenstates of the respective $PT$, $CPT$ or $S$ operators. It is important, however, that the PT and CPT symmetries imposed on the solutions, rule out the phase invariance which is an attribute of the GPEs, because $e^{i\alpha}$ with a real $\alpha$ is not commutative with the $T$ operator. This  artificial constraint on the solutions of the GPEs can be avoided by generalizing the definition of the symmetric modes. Namely, we define a PT [CPT] symmetric solution, $\bpsi^{PT}(x)$ [$\bpsi^{CPT}(x)$], as solutions for which there exist a constant phase  $\alpha$  transforming it into eigenfunctions of the respective symmetry operator, i.e. ensuring $e^{i\alpha}\bpsi^{PT}(x)=PT(e^{i\alpha}\bpsi^{PT}(x))$  [$e^{i\alpha}\bpsi^{CPT}(\tx)=CPT(e^{i\alpha}\bpsi^{CPT}(\tx))$].

Now we turn to a possibility of bifurcations of nonlinear modes obeying the above symmetries from the respective linear Bloch spinors $\bvarphi_{nk}(x)=$col$(\varphi_{nk}^{(1)}(x),\varphi_{nk}^{(2)}(x))$ which solve the  spectral problem $\cH\varphi_{nk}=\mu_n(k)\varphi_{nk}$, where  $k$ refers to the wave-vector in the reduced Brillouin zone (BZ), while
$n$ indicates the number of the branch corresponding to the given $k$ with $n=1$ being the lowest one. The symmetries described above obviously hold also for the eigenfunctions of $\cH$.

\begin{figure}
\includegraphics[width=\columnwidth]{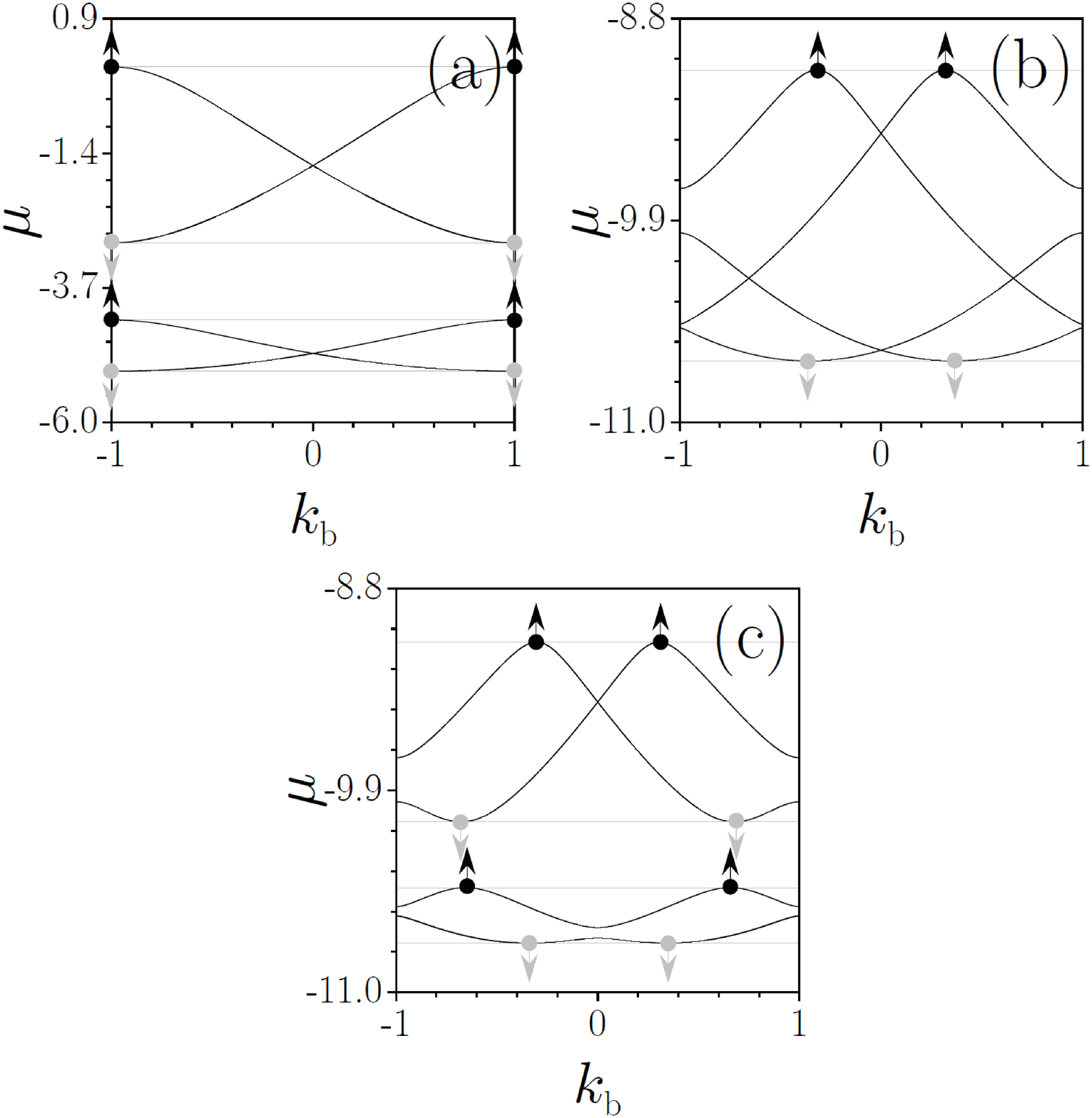}
\caption{(Color online). The spectra of the Hamiltonian $\cH$ with the periodic potential (\ref{omega}) for $\delta=6$ and (a) $\kappa=2$, $\Delta=0$ (b) $\kappa=4.5$, $\Delta=0$,  (c) $\kappa=4.5$, $\Delta=1$. Horizontal gray lines indicate the band-gap edges. Red and blue arrows indicate the points from which gap solitons bifurcate in the case of attractive and repulsive nonlinearities, respectively.
}
\label{fig1}
\end{figure}

In Fig.~\ref{fig1} we show three typical band-gap spectra $\mu_n(k)$. In the absence of the constant splitting ($\Delta=0$) and at relatively weak SO coupling the lower band-edges are reached only at the boundary of the BZ (Fig.~\ref{fig1}a).
Increase of the SO coupling results in closure of the first finite gap and appearance of the crossing points in the spectrum, as well as in a shift of the band-gap edges to some internal points of the BZ (Fig.~\ref{fig1}b), which are symmetric with respect to the center of the BZ and will be denoted as $\pm k_0$.  Finally, in Fig.~\ref{fig1}c we observe "re-opening" of the first gap due to nonzero constant component of the Zeeman field $\Delta$. Notice that the lowest branches of the spectra in Figs. ~\ref{fig1}b and \ref{fig1}c resemble the well known spectrum of the homogeneous SO-BEC.

\begin{figure}
\includegraphics[width=\columnwidth]{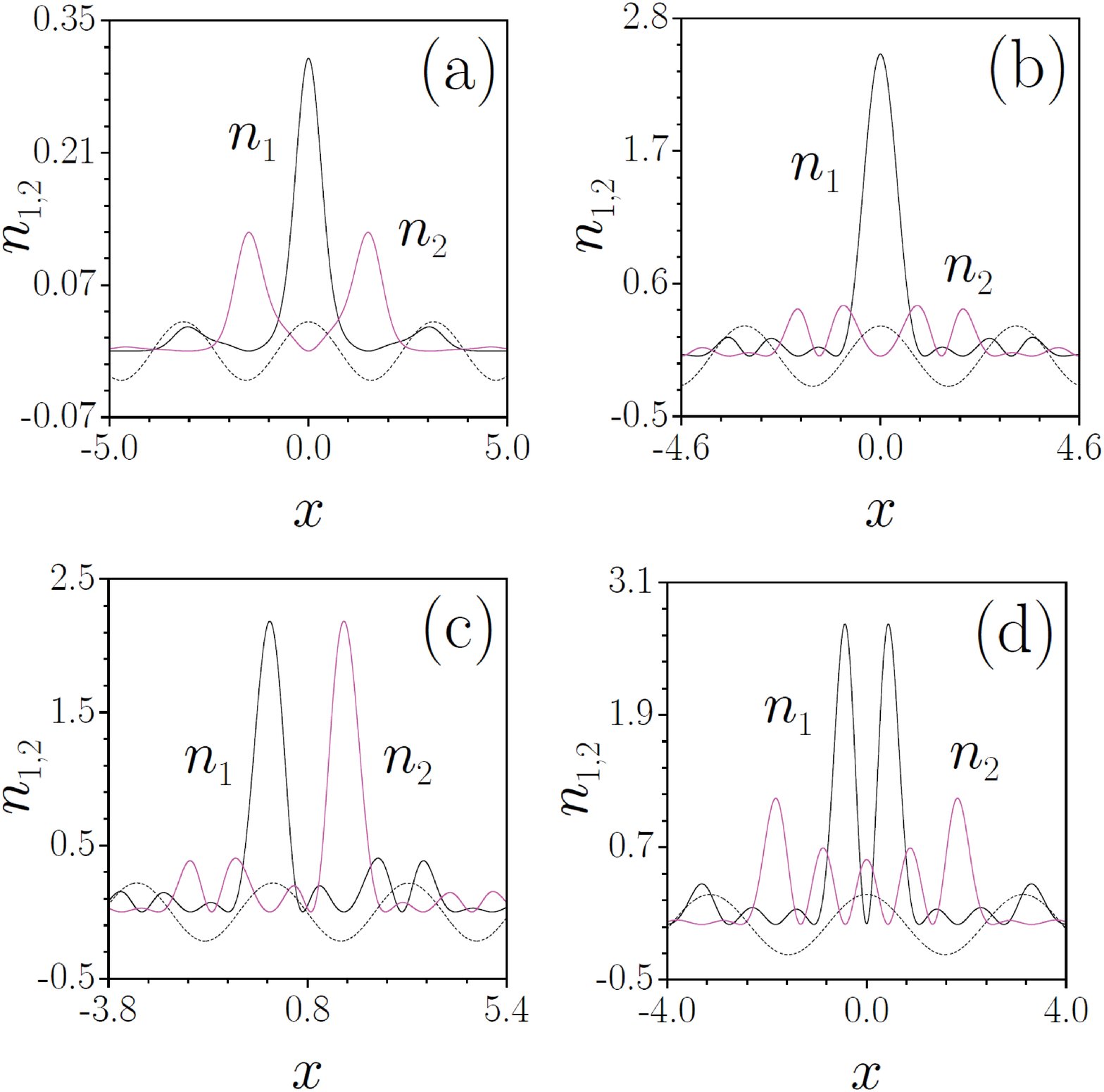}
\caption{(Color online) Densities of a PT-symmetric soliton in the semi-infinite gap at $\mu=-5.23$ \textcolor{black}{obtained} for the attractive nonlinearity (a); PT-symmetric (b) and CPT-symmetric (c) solitons with $\mu=-3$ originating from the first finite gap for the repulsive nonlinearity, and \textcolor{black}{PT-symmetric} soliton with $\mu=-4.03$ from the first finite gap for attractive nonlinearity (d). In all cases $\delta=6$, $\kappa=2$, $\Delta=0$. The dashed line shows the lattice profile for the $|\!\!\uparrow\rangle$ component.}
\label{fig2}
\end{figure}

When the gap edge corresponds to the BZ boundary $k_0=\pm1$ (Fig.~\ref{fig1}a) due to the PT-symmetry the Bloch spinor can be chosen as $\bvarphi_{n1}^{PT}(x)=$col$(v_n^{(1)}(x),iv_{n}^{(2)}(x))$, where $v_{n}^{(1)}(x)= v_{n}^{(1)}(-x)$ and $v_{n}^{(2)}(x)=- v_{n}^{(2)}(-x)$ are real $2\pi-$periodic functions.
 Such Bloch states  give rise to  PT-symmetric gap solitons:  $\psi^{PT}(x)=PT\psi^{PT}(x)$, whose common property is that each spin component has symmetric distribution of the atomic density. Indeed, introducing  $n_j(x)=|\psi^{(j)}(x)|^2$ of the densities of the spinor components one verifies that  $n_j^{PT}(x)=n_j^{PT}(-x)$. For the case of attractive interactions such soliton belonging to the semi-infinite band is illustrated in Fig.~\ref{fig2}a \textcolor{black}{[here and below all soliton solutions of Eq.~(\ref{GPE}) were obtained numerically using the iterative Newton method]}. Notice, that while this situation is seemingly similar to the one observed in the homogeneous medium~\cite{AFKP} the effective periodic potential induced by the Zeeman field is inverted for the different components resulting in  an immiscible anti-ferromagnetic phase (the atoms of each component are located in the minima of the respective periodic potentials). A counterpart of the described solitons can be also obtained in a finite gap in the case of repulsive inter-atomic interactions, see Fig.~\ref{fig2}b. Although there exists a general qualitative similarity with the soliton in the semi-infinite gap (c.f. Fig.~\ref{fig2}a) now   the domains of the atomic concentration (i.e. the points where density is maximal),  especially in the second component, do not coincide with the local minima of the effective potential for the respective component.

Turning to  $CPT$-symmetric solitons, we observe that their characteristic feature  is the equal number of atoms $N_j=\int_{-\infty}^{\infty} n_j(x)dx$ in the both components, $N_1=N_2$, since the densities are now related as: $n_1^{CPT}(\tx)=n_2^{CPT}(-\tx)$.  \textcolor{black}{Since the spinor BEC the components are associated with the opposite spins $+\frac 12$ (the state $|\uparrow\rangle$) and $-\frac 12$ (the state $|\uparrow\rangle$), one can introduce the (pseudo-) magnetization density vectors $M_1=N_{1}/N$ and $M_2=-N_{2}/N$ of the components, such that the total pseudo-magnetization density~\cite{LL8} is given by $M=M_1+M_2
=(N_{1}-N_{2})/{N}
$}. Using this definition we conclude that $CPT$-symmetric solitons have zero total magnetization.  An example of such  a
gap soliton emerging from the first finite gap in a repulsive condensate is presented in Fig.~\ref{fig2}c.
In such modes the maxima of density distributions in the two components are separated by a half of the lattice period and are located in neighboring extrema of the Zeeman lattice $\Omega(x)$.

Finally, in Fig.~\ref{fig2}d we show a PT-symmetric soliton emerging from the first finite gap in the SO-BEC with attractive interactions. Now the symmetries of the spinor components are "inverted" as compared to the case shown in Fig.~\ref{fig2}a, i.e. now $\psi^{(2)}(x)=\psi^{(2)}(-x)$ and $\psi^{(1)}(x)=-\psi^{(1)}(-x)$.
The reason for this "inversion" is that the soliton bifurcates from the upper gap edge (see the upper red arrow in Fig.~\ref{fig1}a), thus resulting in the change of the sign of the effective mass.

\begin{figure}
\includegraphics[width=\columnwidth]{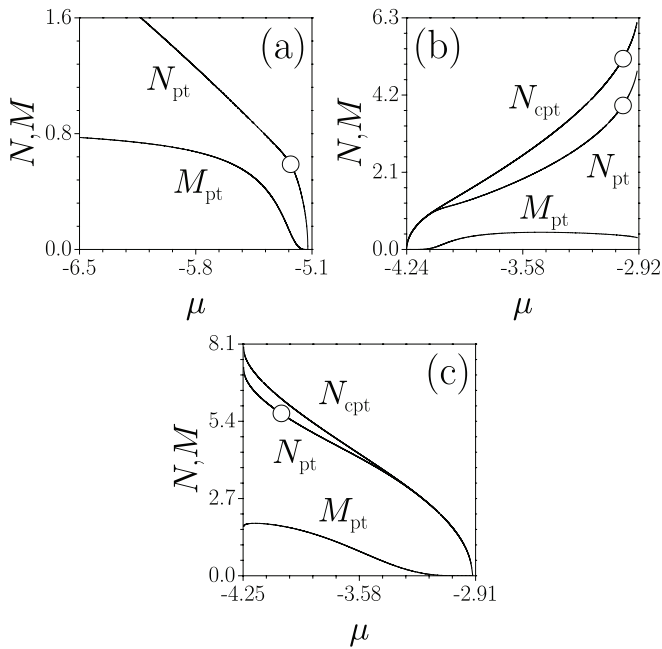}
\caption{The number of particles and magnetization {\it vs} $\mu$ for PT-symmetric solitons emerging from the semi-infinite gap in the case of attractive nonlinearity (a), for PT and CPT symmetric solitons emerging from the first finite gap in the case of repulsive (b) and attractive (c) nonlinearity. In (c) magnetization is multiplied by 10. Circles correspond to solitons shown in Fig.~\ref{fig2}. In all cases $\delta=6$, $\kappa=2$, $\Delta=0$.}
\label{fig3}
\end{figure}

The total numbers of atoms and magnetizations
as functions of the chemical potential are shown in Fig.~\ref{fig3}. For all solitons $N$ demonstrates the typical behavior known for the one-component small-amplitude gap solitons~\cite{rev_mean}: near $n$-th gap edge $N\sim \epsilon$ where   $\epsilon=\sqrt{|\mu_n(k_0)-\mu|/|\mu|}\ll 1$. This indicates that the envelopes $A(\epsilon^2t, \epsilon x)$ of all bifurcating solitons in the small amplitude limit, $\bpsi(x)\approx \epsilon A(\epsilon^2t, \epsilon x)\bvarphi_{nk_0}(x)e^{-i\mu_n(k_0) t}$,  are described by the nonlinear Schr\"odinger (NLS) equation. In the panels b and c one observes that the branches of PT and CPT symmetric solitons seemingly merge. This however occurs only on the plane ($N,\mu$) and do not represent a bifurcation. The both types of the solutions bifurcate from the different linear modes (obeying the respective symmetries). Coincidence of the branches  close to the gap edge is explained by the $S$-symmetry ensuring the equality of the spatially averaged densities of the two components in the linear Bloch spinors, from which the respective solitons bifurcate. CP and CPT solitons in the first finite gap have also different stability~\cite{stability}. In the case of repulsive interactions PT solitons are stable (unstable) at lower (larger) amplitudes. CPT solitons are unstable, while in the vicinity of the bifurcation point they can be classified as metastable ones as the instability increment is small enough ($ <0.03$) within the region of width $|\mu_3(1)-\mu|\approx  0.3$  adjacent to the upper gap edge located at $\mu_3(1)\approx-4.24$.
In the condensate with negative scattering length the CPT-symmetric solitons are always unstable while in the vicinity of the band-edge, i.e. in the small-amplitude limit (with the width of the metastable domain   $|\mu_3(1)-\mu|\approx 0.3$) the PT-symmetric solitons can are metastable.

The magnetization shows rather complex behavior. For sufficiently small detuning $\epsilon$  one observes the behavior approximated by the law $M\sim \epsilon^4$ for all types of the reported PT-symmetric solutions (recall that for the CPT-solitons, like the one shown in Fig.~\ref{fig2}c, the total magnetization is zero). As above, this is explained by the symmetry $S$,  ensuring that at small amplitudes the states $|\!\!\uparrow\rangle$ and  $|\!\!\downarrow \rangle$ are equally populated.  At large detuning the magnetization may start to decay (see Fig.~\ref{fig2}b).

\begin{figure}
\vspace{0.5cm}
\includegraphics[width=\columnwidth]{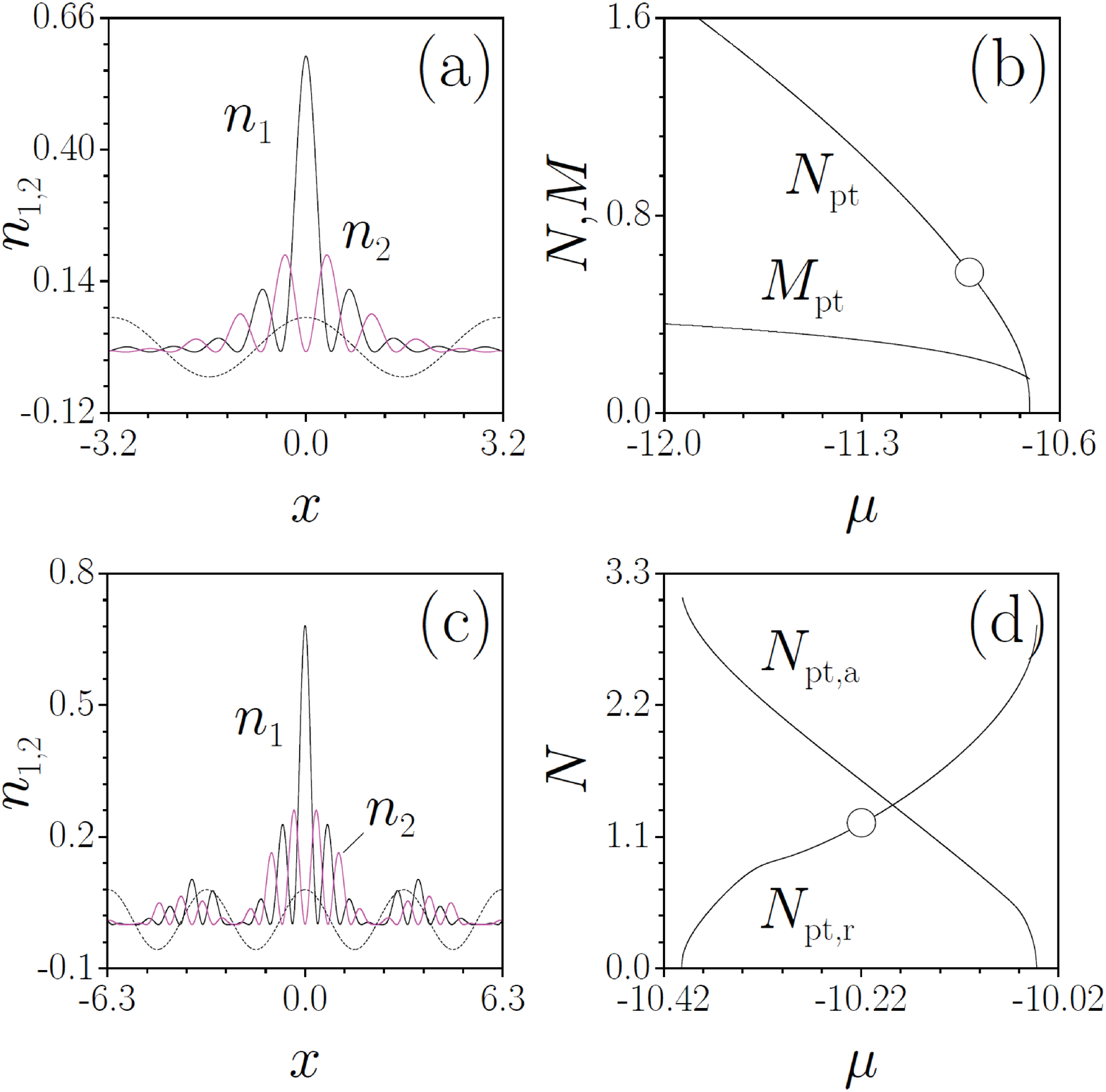}
\caption{(Color online) (a) A gap-stripe soliton from the semi-infinite gap at
$\mu =-10.92$ obtained for attractive nonlinearity. (b) The number of particles and magnetization {\it vs} $\mu$ in the attractive condensate. (c) A gap-stripe  soliton from the first finite gap at $\mu=-10.22$ for repulsive nonlinearity corresponding to the circle in panel (d), which shows $N$ {\it. vs} $\mu$ for repulsive and attractive interactions in the first finite gap. In all cases $\delta=6$, $\kappa=4.5$, $\Delta=1$.}
\label{fig4}
\end{figure}

Now we turn to the situation where a band-gap is achieved in some internal points $\pm k_0$ corresponding to a double degenerate eigenvalue $\mu_n^{(0)}=\mu_n(\pm k_0)$ (see Fig.~\ref{fig1}b and ~\ref{fig1}c).  In the homogeneous case similar situation gives origin to a stripe-phase~\cite{Ho} in repulsive condensate or to a stripe-soliton~\cite{AFKP} in a condensate with a negative scattering length.  In order to construct a PT-symmetric solution we depart from one of the superposition of the linear spinors, say $\bvarphi_{0}(x)=\bvarphi_{nk_0}(x)+\bvarphi_{n,-k_0}(x)$, where the relation between the Bloch states is given by $\bvarphi_{n,-k_0}(x)=\sigma_3 \bar{\bvarphi}_{nk_0}(x)$. PT-symmetric gap-stripe solitons belonging to the branch bifurcating from $\bvarphi_{0}(x)$ are illustrated in Fig.~\ref{fig4} for  the lattice whose band-gap structure is presented in Fig.~\ref{fig1}c. Gap-stripe soliton in the semi-infinite gap (upper row of Fig.~\ref{fig4}) are stable in the whole domain of the existence. In the finite gap the solitons are generically unstable for the depth of potential considered here, although we have observed (numerically) stable time evolutions of  the gap-stripe solitons in the case of repulsive interactions which was accompanied by  regular oscillations of the amplitude, indicating on a possibility of a breather solution. Comparing the gap-stripe solitons with the "conventional" gap solitons considered above, we find that while the envelope of stripe-solitons is also described by the NLS equation, i.e. $N\sim\epsilon$ close to the gap edge, the total magnetization now displays very different behavior near the bifurcation point: it approaches a nonzero value. This difference with the cases of the PT-symmetric solitons follows from the structure of the linear Bloch spinors, which now do not obey S-symmetry and thus the states $|\!\!\uparrow\rangle$ and $|\!\!\downarrow\rangle$ have different populations. In other words such states have nonzero magnetization even in the linear limit.

To conclude we have found a diversity of stable gap and gap-stripe solitons in spin-orbit coupled Bose-Einstein condensate subject to a spatially periodic Zeeman field. The obtained solutions  can be classified by the major physical symmetries, as being invariant with respect to either PT or CPT reductions. The respective solitons have either nonzero or zero total magnetization. The conventional gap and gap-stripe  solitons are observed in lattices with different parameters, while solitons of the same type but obeying different symmetries can be observed in the same lattice but at different spatial locations. All the obtained solutions obey an anti-ferromagnetic structure and thus represent immiscible phases.

\medskip

VVK acknowledges support of the FCT (Portugal) grant PEst-OE/FIS/UI0618/2011.
FKA acknowledges  support from  the Funda\c{c}\~ao Amparo e Pesquisa de S\~ao Paulo(FAPESP, Brasil).

\section{Suplementary Materials to the article}
The stability of a soliton solution, which we denote here as
\begin{eqnarray}
\bPsi^{sol}(x)e^{-i\mu t}=[\bU(x)+i\bV(x)]e^{-i\mu t}
\end{eqnarray}
 splitting the real and imaginary parts, i.e. $\bU=$Re$(\bPsi^{sol})$ and $\bU=$Re$(\bPsi^{sol})$, was established using the linear stability analysis. To this end, following the standard algorithm, we make the ansatz ($j=1,2$)
\begin{eqnarray}
\label{subs}
\bPsi (x,t)=[\bPsi^{sol}(x)+\bp(x,t)]
e^{-i\mu t}
\end{eqnarray}
where
$$
\bp(x,t)=\left(
\begin{array}{c}
p_1(x,t)
\\
p_2(x,t)
\end{array}
\right)=o(\bPsi^{sol}(x))
$$ is a small perturbation in the coupled Gross-Pitaevskii equations
\begin{equation}
i\bPsi_t = -\frac{1}{2}\bPsi_{xx} + \frac{\sigma_3}{2}\Omega(x)\bPsi  +
i\kappa\sigma_1 \bPsi_x
+ g(\bPsi^\dag\bPsi)\bPsi.
\label{GPE}
\end{equation}
and linearize them with respect to small $\bp(x,t)$. This gives us the evolution problem for the perturbation $\bp(x,t)$ which in the matrix form reads
\begin{eqnarray}
\label{linear1}
\frac{\partial}{\partial t}\bP=\hat{L}\bP, \quad
\bP=\left(
\begin{array}{c}
\mbox{Re}\, p_1(x)
\\
\mbox{Im}\, p_1(x)
\\
\mbox{Re}\, p_2(x)
\\
\mbox{Im}\, p_2(x)
\end{array}
\right)
\end{eqnarray}
with the matrix operator $\hat{L}$ given by
\begin{widetext}
\begin{eqnarray}
\label{linear2}
\hat{L}=\left(
\begin{array}{cccc}
2gU_1V_1 & L_-+2gV_1^2&  \kappa \frac{\partial}{\partial x}+2gU_2V_1  & 2gV_1V_2
\\
-L_--2gU_1^2 & -2gg U_1V_1& -2gU_1U_2 & \kappa \frac{\partial}{\partial x}-2gU_1V_2
\\
\kappa \frac{\partial}{\partial x}+2gU_1V_2 & 2gV_1V_2 & 2gU_2V_2 & L_++2gV_2^2
\\
-2gU_1U_2 & \kappa \frac{\partial}{\partial x}-2gU_2V_1 & -L_+- 2U_2^2 & -2gU_2V_2
\end{array}
\right)
\end{eqnarray}
 where the operators $L_\pm$ are defined by
\begin{eqnarray}
\label{linear3}
L_\pm&=&-\frac 12 \frac{\partial^2}{\partial x^2}  + g(|\Psi_1^{sol}(x)|^2+|\Psi_2^{sol}(x)|^2) \pm \frac{\Omega(x)}{2}-\mu
\nonumber
\\
&=&-\frac 12 \frac{\partial^2}{\partial x^2}  + g[U_1^2(x)+V_1^2(x)+U_2^2(x)+V_2^2(x)] \pm \frac{\Omega(x)}{2}-\mu
\end{eqnarray}
\end{widetext}
Finally, we look for a solution of (\ref{linear1}) -- (\ref{linear3}) in the form $\bP(x,t)=\tilde{\bP}(x)e^{\lambda t}$ where $\lambda$ is a constant spectral parameter and $\tilde{\bP}(x)$ is $4\times 1$-column time-independent vector, what gives us the linear spectral problem:
\begin{eqnarray}
\label{linear4}
\lambda\tilde{\bP}=\hat{L}\tilde{\bP}
\end{eqnarray}
The eigenvalue problem (\ref{linear4}) was solved numerically to obtain the dependence of the growth rate $\lambda$ on the chemical potential $\mu$. The soliton is unstable if the respective $\lambda$ has a real positive part.


\begin{thebibliography}{99}
\bibitem{com} The Wigner time inversion operator $T\bPsi(t,x)= \bar{\bPsi}(-t,x)$ applied to a stationary  mode $\bPsi(t,x)=e^{i\mu t}\bpsi(x)$ is reduced to the complex conjugation.
\end{thebibliography}
\end{document}